\documentclass[epsf,twocolumn,preprintnumbers]{revtex4}
\usepackage{epsfig,graphics}% Include figure files
\usepackage{dcolumn}% Align table columns on decimal point
\usepackage{bm}% bold math
\usepackage{amsmath}
\usepackage{xcolor}
\usepackage{comment}

\definecolor{dkred}{rgb}{0.6,0.0,0.0}
\definecolor{dkblue}{rgb}{0.123,0.247,0.554}

\begin{document}

%\title{DFT+U+"Exact Diagonalization" (ED) for  UTe$_2$}
%\title{Correlated Quasiparticle Dispersion and ARPES Spectra in UTe$_2$}

\title{UTe$_2$: a nearly insulating half-filled 
        $j=\frac{5}{2}$  $5f^3$ heavy fermion metal \\
  }

\author{Alexander  B.  Shick}
\affiliation{Institute of Physics, Czech Academy of Science,  Na Slovance 2, CZ-18221 Prague, Czech Republic}
\email{shick@fzu.cz}
\author{Shin-ichi Fujimori}
\affiliation{Materials Sciences Research Center, Japan Atomic Energy Agency, Sayo, Hyogo 679-5148, Japan}
\author{Warren E. Pickett$^*$}
\affiliation{Department of Physics, University of California Davis, Davis CA 95616} 
\email{pickett@physics.ucdavis.edu}
\date{\today}
%\pacs{}
\date{\today}

\begin{abstract}
Correlated band theory implemented as a combination of  
density functional theory with exact diagonalization [DFT+U(ED)] of the Anderson
impurity term with Coulomb repulsion $U$ in the open 14-orbital $5f$ shell 
is applied to UTe$_2$.  
The small gap for $U$=0, evidence of the half-filled $j=\frac{5}{2}$
subshell of $5f^3$ uranium, is converted for $U$=3 eV to a flat band 
semimetal with small heavy-carrier Fermi surfaces that will make properties sensitive to
pressure, magnetic field, and off-stoichiometry, as observed experimentally.
Two means of identification from the Green's function give a mass enhancement 
of the order of 12 for already heavy (flat) bands,
consistent with the common heavy fermion characterization of UTe$_2$. The
predicted Kondo temperature around 100 K matches the experimental values
from resistivity.
The electric field gradients for the two Te sites are calculated by
DFT+U(ED) to differ by a factor of seven, indicating a strong site distinction,
while the anisotropy factor $\eta=0.18$ is similar for all three sites. 
The calculated uranium moment $<M^2>^{1/2}$ of
3.5$\mu_B$ is roughly consistent with the published experimental Curie-Weiss values
of 2.8$\mu_B$ and 3.3$\mu_B$ (which are field-direction dependent), 
and the calculated separate spin and orbital
moments are remarkably similar to Hund's rule values for an $f^3$ ion.
The $U$=3 eV spectral density is compared
with angle-integrated and angle-resolved photoemission spectra, with 
agreement that there is strong
$5f$ character at, and for several hundred meV below, the Fermi energy.
Our results support the picture that the underlying ground state of UTe$_2$ 
is that of a half-filled $j=\frac{5}{2}$ subshell with two half-filled
$m_j=\pm\frac{1}{2}$ orbitals forming a narrow gap by hybridization, then
driven to a conducting state by configuration mixing (spin-charge fluctuations).  
UTe$_2$ displays similarities to UPt$_3$ with its $5f$ dominated Fermi
surfaces rather than a strongly localized Kondo lattice system.
\end{abstract}
%\pacs{71.20,71.27+a,75.40.Cx} 

\maketitle

\section{Introduction}

Recently discovered superconductivity (SC) in the heavy fermion material 
UTe$_2$~\cite{Ran2019,Aoki2019} below 1.7K, earlier studied in single
crystal form,\cite{Ikeda2006} shows a number of peculiar
aspects. Unlike in several other U superconductors (UGe$_2$, UCoGe, 
URhGe, UCoAl)
which display coexisting superconductivity and ferromagnetism,
no long range magnetic order in the ground state has been observed. The magnetic
susceptibility~\cite{Ran2019,Aoki2019} has Curie-Weiss character, with magnetic moment
in 2.8-3.3$\mu_B$/U range. Magnetic moments in a metal that do
not order suggest UTe$_2$ is associated with a class of conducting spin liquids.
The large and anisotropic Curie-Weiss magnetic susceptibility suggest antiferromagnetic
coupling, whereas other probes [NMR Knight shift and spin-lattice relaxation rate ($1/TT_1$)] suggest 
critical ferromagnetic fluctuations that could mediate SC in 
UTe$_2$.\cite{Ran2019,Aoki2019,Tokunaga2019} The phase diagrams versus field and its
direction, temperature, and pressure -- including re-entrant superconductivity with
application of field -- are unusually complex, but those complications
will not be addressed in this paper.

\begin{figure}
%\vskip 8mm
\centerline{\includegraphics[width=1.0\columnwidth]{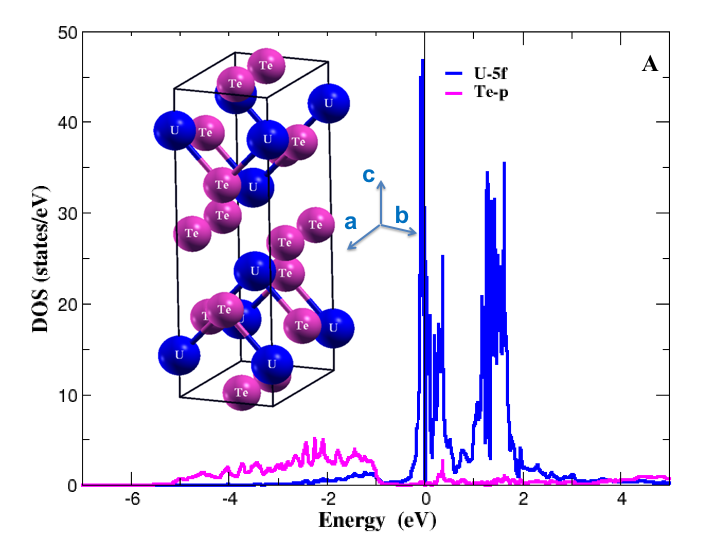}}
\centerline{\includegraphics[width=1.0\columnwidth]{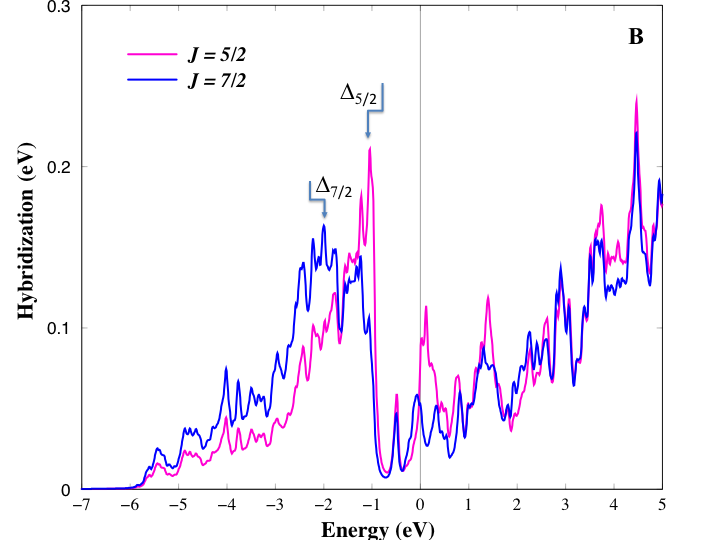}}
\centerline{\hspace*{-0.75 cm} \includegraphics[width=0.95\columnwidth]{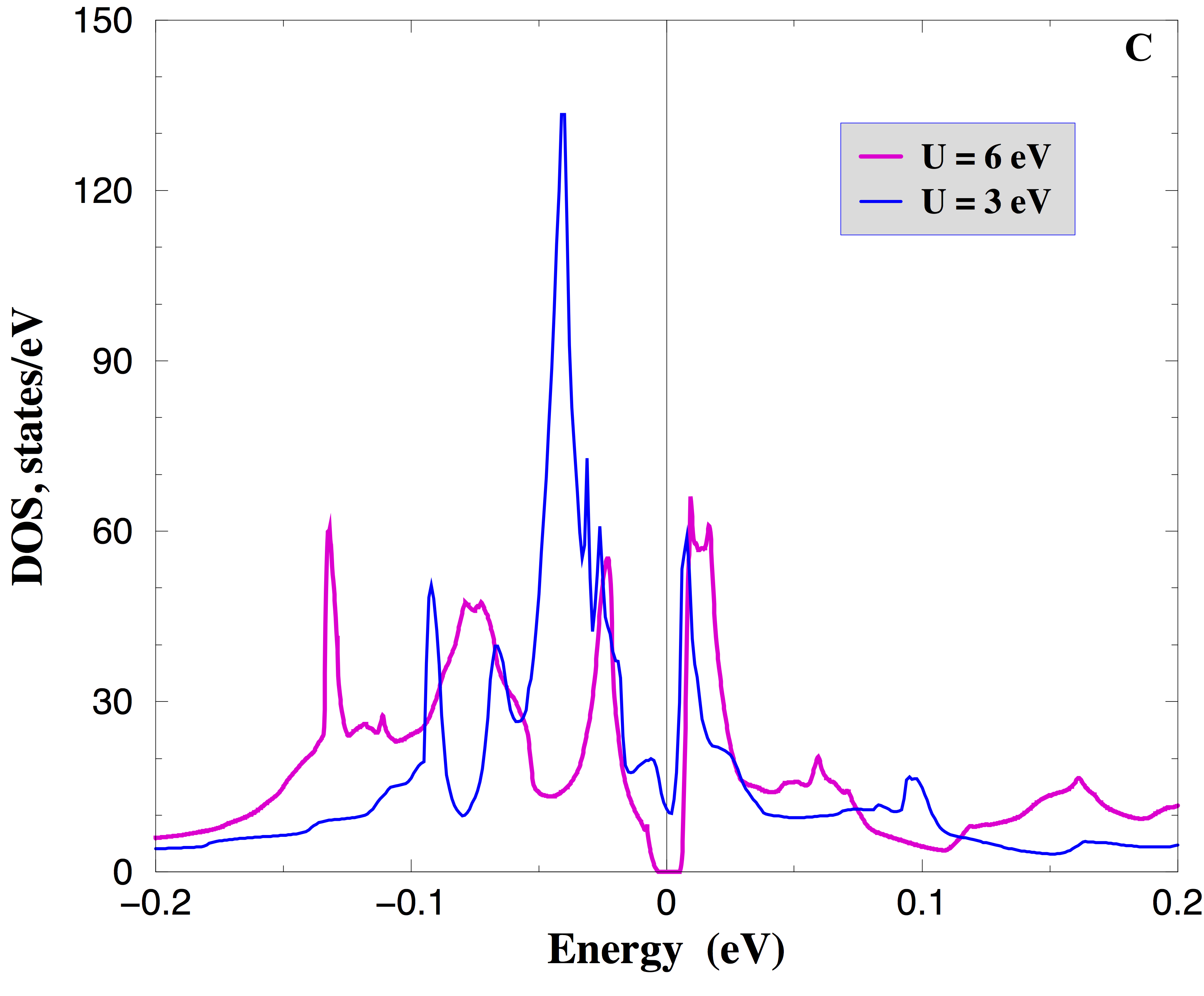}}
\caption{(Color online)
(A) Projected densities of states/eV (per unit cell) for nonmagnetic
UTe$_2$ from the DFT+U(OP) functional ($U$=$J$=0.5 eV), showing the 
projections for U $5f$ (blue) and
Te $5p$ (magenta).
Inset: the orthorhombic $Immm$ UTe$_2$ crystal structure.
(B) The $j=\frac{5}{2}$ and $\frac{7}{2}$  projected hybridization functions 
${\Delta}(z)={1 \over {\pi n_f}}\mathop{\rm Im}\mathop{\rm Tr} [G_0{}^{-1}(z)]$  (eV)
obtained from LDA calculations~\cite{Shick2019}.
(C) Total densities of states/eV-unit cell,
for nonmagnetic UTe$_2$ from the DFT+U(ED) functional. Note that $U$=6 eV
opens a gap.
}
\label{str-hyb}
\end{figure} 

UTe$_2$ is however a heavy fermion metal, with a large Sommerfeld 
coefficient $\gamma\approx 120$
mJ/K$^2$,\cite{Ran2019,Aoki2019} indicative of Kondo screening of local moments.
The resistivity $\rho\sim 1 m\Omega$ cm is slowly increasing from room temperature 
down to 75K, then it decreases rapidly over two orders of magnitude just above T$_c$, typical
heavy-fermion (HF) behavior with a Kondo temperature of $\sim$100K.\cite{Ran2019} The susceptibility for field along 
the easy $a$ axis increases strongly below 100K, becoming a factor of five or more larger 
than in the other two directions. The interpretation has been that of critical magnetic 
fluctuations around incipient FM order along the $a$ axis. A considerable number of 
measurements of electrical and thermal conductivity, NMR spectra, and penetration depth 
in magnetic fields gives strong justification of a point-node, Weyl superconducting state.   

Based on a non-vanishing specific heat coefficient $C_v/T$ below T$_c$, 
UTe$_2$ had originally been suggested by Ran and 
collaborators~\cite{Ran-arxiv} to provide a new phase of superconducting matter with
a `Bogoliubov Fermi surface,' a Fermi surface that is gapless over not points, or
lines, but an area.
However, subsequent extension of the measurements to
50 mK revealed an {\it upturn} in $C_v/T$. The normal state $C_v$, {\it i.e.} above
the critical magnetic field) was fit with the conventional form augmented by
a residual contribution $C_x$ of unknown origin, modeled by a $C_x/T=A_{div}T^{-1/3}$
form in the region of 50mK to T$_c$.
$C_x$ would arise from thermal excitations
that are not gapped by SC.\cite{Ran2019,Aoki2019,Metz2019}
Supposing this extrapolation is reasonable
up to around 2K (taken literally, it would give a divergent entropy), 
the excess entropy is (we estimate) of the order of $S_x\sim
\alpha_x k_B ln2$, with $\alpha_x\sim 10^{-2}$, small enough to be
extrinsic in origin.

\begin{comment}
below T$_c$=1.7~K in which half of the electrons become superconducting and
half remain  normal (thus with Fermi surfaces), based on heat capacity 
(C$_v$) data  down to 1.5 K. This result was confirmed of by
Aoki {\it et al.}~\cite{Aoki2019}.
%who extended the characterization of UTe$_2$.\cite{Aoki_2,Aoki_3,Aoki_4}
One scenario was that, at T$_c$, some additional symmetry is broken beyond
the usual broken gauge symmetry that produced a new and unusual state of
matter with {\it area} of gap nodes, very different from the known possibilities
of point and line nodes.
%While no magnetic ordering is detected, it was proposed that half of the
%fermionic excitations (spin up, for lack of a better characterization) 
%become superconducting 
%while the spin down fermions remain in the normal conducting state. 
\end{comment}

U-based compounds provide a spectacular variety of behaviors, from ground
states to unusual spectra. The ferromagnetic superconductors mentioned above 
have been reviewed and compared by Aoki, Ishida, and
Floquet.~\cite{Aoki2019a} These compounds seem to require correlation
corrections beyond conventional density functional theory (DFT) in its
local or semilocal approximations to account for their properties. 
Superconducting UPt$_3$ (T$_{c} <$1~K), on the other hand, has six complicated
Fermi surfaces that are described quite well by DFT 
calculations,\cite{Wang1987,McMullan} with 
excellent agreement requiring only
energy shifts of a few tens of meV. Isovalent and
isostructural UPd$_3$, on the other
hand, is reproduced only when the U $5f$ states are localized (removed from
the valence states).\cite{Oguchi1986} 
%{\bf We need references.}  

The electronic structure of UTe$_2$ has been studied from a first-principles
itinerant (local density approximation [LDA]) 
viewpoint making use of density functional theory 
(DFT), initially by Aoki and collaborators~\cite{Aoki2019} and
Fujimori {\it et al.}\cite{Fujimori2019}. LDA however predicts a small gap
for  this heavy fermion metal, so these works were followed by others
who have incorporated some correlation correction related to a (semi)localized
nature of the $5f$ states.
Two of the current authors applied a DFT+orbital 
polarization approach [DFT+U(OP), using a Coulomb repulsion $U$=$J$],
obtaining heavy $5f$ bands giving large Fermi surfaces (FSs).\cite{Shick2019}  
In these calculations the uranium 5$f$ states are treated as basically itinerant, with the
repulsion $U$ and Hund's exchange $J_H$ encouraging orbital polarization. 
Analysis indicated two roughly half-filled $5f$ orbitals, $m_j=\pm\frac{1}{2}$,
at the Fermi level, suggesting multiband half-filled physics.  

Some of the most basic theoretical questions are unsettled. Standard DFT
calculations give a U $5f$ occupation near $f^3$.
From the viewpoint of strongly localized $5f$ orbitals ($U$=6 eV), 
Miao {\it et al.} investigated the effect of removing 
the 5$f$-states from consideration, {\it i.e.} using ThTe$_2$ without $5f$
occupation as an underlying model of Fermi level bands.\cite{Miao2019} 
Applying dynamical mean field theory (DFT+DMFT)
to UTe$_2$ with a large value of repulsion $U$ (with similar results obtained
by Xu {\it et al.}\cite{Xu2019}), the $5f$ bands were shifted in Mott insulator fashion
away from E$_F$, leaving a dispersive Te band and a large Fermi surface that
was supported by their angle-resolved photoemission spectra (ARPES).
A similar shift of $5f$ bands was observed by 
Ishizuka and coauthors\cite{Ishizuka2020} applying
the DFT+$U$ method, even for moderate values of $U$ in the 1-2 eV range.
ARPES data taken at higher energy with longer escape depth however give
clear evidence of strong $5f$ character extending to E$_F$.\cite{Fujimori2019}
Both the DFT+$U$ and DFT+DMFT methods predict a predominant $5f^2$ 
configuration (our own DFT+$U$ calculations do so as well),
whereas uranium core level spectroscopy\cite{Fujimori2020} indicates the $5f^3$
configuration is dominant, as in several other itinerant uranium intermetallics. 
We will present evidence that supports a dominant $f^3$ configuration.
We address these differences within this paper, with some emphasis on
angle-integrated PES (AIPES).

Less strongly correlated methods, such as we will present, leave 
flat $5f$ bands at E$_F$ in a hybridized band picture consistent with $5f$
dominated Fermi surfaces and with AIPES
data, whereas transport and thermodynamic properties will be enhanced by
residual (dynamic) interactions. Note that some U compounds, 
for example UPt$_3$ and UBe$_{13}$,
are also heavy fermion metals that do not order magnetically but display
exotic superconducting gap symmetry and $5f$ electron dominated Fermi 
surfaces.\cite{Aoki2019a} 

This itinerant-localized dichotomy is itself not so unusual in metallic actinide 
compounds, reflecting the ``dual" nature~\cite{Zwicknagl2007} often 
exhibited by open 5$f$ shells. Early on, Hill recognized the shortest U-U
separation as a critical parameter in uranium intermetallics and provided the
``Hill plot''\cite{Hill1970,Boring2000} of ordering temperature 
(magnetic or superconducting) versus
U-U separation. If shorter
than 3.5~\AA, the $5f$ states are itinerant and sometimes superconducting, 
if longer they display a local moment and usually order. 
This critical separation is not absolute: the
Hill criterion was found to be violated when the heavy fermion metals
UPt$_3$ and UBe$_{13}$ were synthesized and found to be superconducting
without magnetic order. Still, the Hill criterion is a very useful and
physically motivated guide. 
In this picture, the nearest neighbor U-U separation in UTe$_2$ of 3.78~\AA~
puts it well into the localized regime of the Hill plot: magnetic rather than
superconducting. Just how UTe$_2$ violates the Hill criterion is a fundamental
question in the understanding of this fascinating compound.
  
This dual nature of the $5f$ shell requires that both local and itinerant features of the 
$f$-electrons may need to be allowed in a description of the electronic
structure of UTe$_2$.
Ishizuka {\em et. al.}~\cite{Ishizuka2020} performed DFT+$U$ calculations 
(related somewhat to the DFT+$U$(OP) results mentioned above)
for empirically reasonable values of the effective Coulomb $U_{eff}$ = 
$U-J$ = 1.1--2.0 eV, $J$=0.
Constraining their DFT+$U$ calculations to the nonmagnetic phase, 
they obtained a metallic band structure, with reconstruction occurring around
$U\sim$1.1 eV.  Contrary the results of Refs.~[\onlinecite{Miao2019,Xu2019}], 
the electronic states near the Fermi energy ($E_F$) formed narrow bands with 
predominantly 5$f$-character. The fundamental limitation of these calculations 
is that they rely on a single-determinant approximation for the 5$f$-manifold 
of the U atom, and the DFT+$U$ method is more adept in describing 
occurrence and effects of long-range magnetic order. 
Our calculations parallel to those of Ishizuka {\it et al.}
indicate that their choice of Hund's exchange $J$=0, which neglects the
anisotropy $U_{m,m'}$ of the repulsive interaction and misses Hund's 
exchange, has a substantial effect of the resulting band structure.

\begin{comment}
Such a magnetic subsystem may exhibit behavior characteristic of a spin
liquid\cite{Savary2017} or that of a spin glass.\cite{Mydosh2015}
YMn$_2$ and CaCo$_{1.86}$As$_2$ both are magnetic metals that have
been discussed as spin liquids,\cite{Nakamura1997, Sapkota2017} but unlike
Eu they are understood in terms of frustrating short-range interactions.
however seem as fundamental as the difference in ground states of the undoped materials:
antiferromagnetic (AFM) insulator for CCO, and disordered moment 
conductor for NNO, the latter
being a conducting quantum paramagnetic (MQPM) phase in the Sachdev-Read
classification\cite{sachdev}.
This correlated spin liquid phase,
the Sachdev-Read MQPM phase that has been proposed in another disordered
moment superconductor\cite{stpi2019},
provides the platform for the superconductivity that appears upon hole doping.
\end{comment}

In this work we study electron correlation effects in UTe$_2$ beyond the 
static mean-field DFT+$U$ approximation. We extend DFT method making use 
an Anderson impurity treatment of the $5f$ shell treated with
exact diagonalization (ED) techniques and including the full self-consistency 
over the charge density. In Sec.~II %%%and Appendix A 
we describe the basic 
equations of the DFT+$U$(ED) method. In Sec.~III the electronic structure results 
for a conventional value of Coulomb $U$=3 eV are presented and compared to large 
Coulomb $U$=6 eV results. Sec. IV is devoted to a comparison with 
photoemission results,
and a Summary in Sec. V concludes the paper.

\section{Structure and Methods}
\subsection{Structure}
UTe$_2$ crystallizes in a body-centered orthorhombic $Immm$ structure~\cite{Ikeda2006}
(space group \#71) with 
$a$=4.161~\AA, 
$b$=6.122~\AA, 
$c$=13.955~\AA, this volume
containing two formula units (f.u.) (see~Fig.\ref{str-hyb}A). 
The atomic sites and symmetries are:
 U, (0,0,0.13544) $4i$ $mm2$; 
Te1, ($\frac{1}{2},0,0.2975)$  $4j$ $mm2$; 
Te2, (0,0.2509,$\frac{1}{2}$) $4h$ $m2m$. 
The U-U separations are, in increasing order, 
3.78~\AA~dimer separations directed along the $c$-axis, 
4.16~\AA~along the $a$-axis (the lattice constant), 
4.89~\AA~in the $b-c$ plane, and 
6.12~\AA~along the $b$-axis (the lattice constant). The normal to
the cleavage plane lies 23.7$^{\circ}$ from the $b$-axis in the
$b-c$ plane. The structure is sometimes pictured as U ladders
lying in the cleavage plane, or even two sets of U ladders, but the large
U-U separation makes any quasi-one-dimensional aspects difficult to identify.
Note: much of our description will use the orthorhombic pseudo-zone with
dimensions $\frac{2\pi}{a}, \frac{2\pi}{b}, \frac{2\pi}{c}.$ 

\subsection{Formalism of the DFT+U(ED) method}
Given the combination of itinerant behavior with localized physics in actinide
materials and the Fermi liquid character of UTe$_2$, 
we aim for an effective (low energy) band structure that nevertheless
includes essential effects of on-site repulsion and Hund's exchange on the
uranium site and related configurations, and does so self-consistently. 
We use an extension of the widely used DFT+U method\cite{ylvisaker} 
that makes use of a combination
of DFT with the exact diagonalization (ED) of the multiconfigurational $5f$
shell of uranium, in the spirit of a generalized and orbital-occupation
(charge, with spin) self-consistent generalization of the Anderson impurity model
analogous that that done in dynamical mean field theory. 

This DFT+U(ED) method takes advantage of the fact that electron interactions in
the $s$, $p$, and $d$ shells are well described in DFT, whereas 
interelectronic interactions within the $5f$ shell are treated explicitly. 
We use the fully anisotropic, rotationally invariant implementation of the 
$+U$ interaction in the DFT+U method in the full-potential 
linearized augmented plane wave 
(FP-LAPW) basis that includes both scalar-relativistic and spin-orbit 
coupling (SOC) effects~\cite{shick99,shick01}.
The calculations were carried out in the observed paramagnetic state,
however both spin and orbital moment effects, such as rms
values of the moments, are included in this 
U(ED) extension.

The effects of the interaction Hamiltonian
$H_{\rm int}$ on the electronic structure are described with
the aid of an auxiliary  impurity model  describing the complete
fourteen-orbital U 5$f$ shell. This multi-orbital  impurity model  includes 
the full Coulomb interaction, the SOC that is very
strong in both U and Te, and the crystal
field. The corresponding Hamiltonian can be written (see, for example,
Hewson\cite{Hewson}),
\begin{align}
\label{eq:hamilt}
H_{\rm int}  = & \sum_{\substack {q m m' \\ \sigma \sigma'}}
 \epsilon^{q}_{m\sigma,m'\sigma'} b^{\dagger}_{qm\sigma}b_{qm'\sigma'}
 +\sum_{m\sigma} \epsilon_f f^{\dagger}_{m \sigma}f_{m \sigma} 
\nonumber \\
& + \sum_{mm'\sigma\sigma'} \bigl(\xi {\bf l}\cdot{\bf s}
  + \Delta_{\rm CF}\bigr)_{m\sigma,m'\sigma'}
  f_{m \sigma}^{\dagger}f_{m' \sigma'}
\nonumber \\
& +  \sum_{\substack {q m m' \\ \sigma \sigma'}}   \Bigl(
   V^{q}_{m\sigma,m'\sigma'}
 f^{\dagger}_{m\sigma} b_{qm' \sigma'} + \text{h.c.}
  \Bigr)
\\
& + \frac{1}{2} \sum_{\substack {m m' m''\\  m''' \sigma \sigma'}}
  U_{m m' m'' m'''} f^{\dagger}_{m\sigma} f^{\dagger}_{m' \sigma'}
  f_{m'''\sigma'} f_{m'' \sigma}.
\nonumber
\end{align}
Here $f^{\dagger}_{m \sigma}$ creates an electron in orbital $m$
and spin $\sigma$ in the 5$f$ shell
and $b^{\dagger}_{m\sigma}$ creates an electron in state $m\sigma$
in the ``bath'' that
consists of those host band states that hybridize with the impurity
5$f$ shell. 
$\epsilon_f$ is the energy position of the non-interacting 
$5f$ `impurity' level,
and $\epsilon^{q}$ are the bath energies. 

The parameter $\xi$ specifies the strength of
SOC obtained from the atomic potential, and $\Delta_{\rm CF}$ is 
the crystal-field (CF) potential at the
impurity, as described below. The matrices  $V^{q}$ 
describe the hybridization
between the 5$f$ states and the bath orbitals at energy
$\epsilon^{q}$. 

In these calculations two sets of Slater integrals (in eV)\\
$F_0=3.00$, $F_2=6.024$, $F_4=4.025$, $F_6=$2.94, and\\
$F_0=6.00$, $F_2=6.024$, $F_4=4.025$, $F_6=$2.94 \\
were chosen
to specify the Coulomb interaction matrix $U_{m m' m'' m'''}$
in Eq.~(\ref{eq:hamilt}).
They correspond to values of $U=3$~eV and $U=6$~eV respectively,
with exchange~$J=0.51$ eV.

\subsection{Computational procedure}
\subsubsection{Overview}
The calculation uses the same DFT foundation and interaction matrix as 
implemented
by Shick and collaborators\cite{shick99}  and used in conventional
DFT+U calculations, but it generalizes the Shick {\it et al.}\cite{shick09} ionic
limit approximation (``Hubbard I'') to 
(a) include hybridization with the
environment, and (b) applies 
exact diagonalization to evaluate the impurity Green's function, much
like some versions of dynamical mean field theory. A
crystal field potential is included in the formalism, but for UTe$_2$ it is
expected that the DFT treatment of CF is sufficient, as done for other
actinides.\cite{shick09} 

To specify the bath parameters, we
use the previously reported LDA results for 
non-magnetic UTe$_2$~\cite{Shick2019} (the conventional basic
underlying electronic structure before many-body interaction is
considered), repeated for this study. Since
the interaction $U$-terms in $H^{int}$ are couched in spin-orbital rather than
$j, j_z$ language, it is judicious to treat SOC in $H_{int}$.
It is assumed that the first and fourth terms in
Eq.~(\ref{eq:hamilt}) are diagonal in $\{j,j_z\}$ representation.

Next, we obtain $V_{q=1}^{j}$ and $\epsilon_{q=1}^{j}$,
for $j=\frac{5}{2}$ and $\frac{7}{2}$,
from the hybridization functions 
\begin{eqnarray}
{\Delta}_j(z) = {1 \over {\pi n_j}} \mathop{\rm  Im}\mathop{\rm Tr_j}
[G_0{}^{-1}(z)] 
\end{eqnarray}
where $Tr_j$ is the trace over the $j$-subspace, with $n_f=6$ for $j=5/2$,
$n_f=8$ for $j=7/2$.  
$G_0(z)$ is the non-interacting DFT Green's function
extended to the complex energy $z$ plane, equal to the crystal Green's
function in Eq.~(\ref{eq:gf}) with the self-energy
$\Sigma$ set to zero. The hybridization functions ${\Delta_j}(\epsilon)$,
obtained from the LDA calculation without additional adjustment,
 are shown in Fig.~\ref{str-hyb}B.
Since the essential hybridization occurs in the energy region of Te $p$ 
states [see Fig.~\ref{str-hyb}(A)], we set 
$\epsilon_{q=1}^{5/2}$ to the $-1.045$ eV peak position of 
  ${\Delta}^{5/2}(\epsilon)$, and  
$\epsilon_{q=1}^{7/2}$ to the $-1.996$ eV peak position of 
  ${\Delta}^{7/2}(\epsilon)$. From the value of 
$\Delta(\epsilon_{q=1})=V^2 \delta(\epsilon-\epsilon_{q=1})$,  
  we obtain $V_{q=1}^{5/2}=0.459$ eV and $V_{q=1}^{7/2}=0.404$ eV. 

\subsubsection{Local approximation for $\Sigma(z)$}
The band Lanczos method~\cite{J.Kolorenc2012} is employed to find
the lowest-lying eigenstates of the many-body Hamiltonian $H_{\rm
imp}$ and to calculate the one-particle Green's matrix $[G_{\rm
imp}(z)]_{\gamma,\gamma'}$ in the subspace of the $f$ spin-orbitals 
$\{\phi_{\gamma} =  \phi_{m \sigma} \}$
at low temperature ($k_{\rm B}T= \beta^{-1}=(1/500)$ eV, T$\sim $40 K). 
The expression for $G_{\rm imp}$ is 
\begin{eqnarray}
[G_{\rm imp}(z)]_{\gamma\gamma'}=\frac{1}{Z}\sum_{\alpha,\delta}
  \frac{ \langle \alpha|c_{\gamma}|\delta\rangle 
         \langle |\delta|c^{\dag}_{\gamma'}|\alpha\rangle }
       {z+E_{\delta}-E_{\alpha}}
    [e^{-\beta E_{\delta}} + e^{-\beta E_{\alpha}}],
\label{eq:gimp}
\end{eqnarray}
where $Z$ is the partition function, $E_{\alpha}$ is the energy of
the eigenstate $|\alpha\rangle$ of Eq.(\ref{eq:hamilt}), and
$z$ is the (complex) energy.
The self-energy matrix $\Sigma_{\gamma,\gamma'}(z)$ is then
obtained from the inverse of the Green's function matrix
$G_{\rm imp}$.

The self-energy is then inserted into the local Green's function $G(z)$ 
\begin{equation}
G_{\gamma \gamma'}(z) = 
\int_{\rm BZ} \frac{{\rm d}^3 k}{V_{\rm BZ}} \,\bigl[z+\mu-H_{\rm DFT}({\bf
k})-\Sigma(z)\bigr]^{-1}_{\gamma \gamma'}\,, 
\label{eq:gf}
\end{equation}
calculated in a single-site approximation as described 
previously,\cite{shick09,Kristanovski2018} from which 5$f$ orbital occupations 
are obtained. $V_{BZ}$ is the volume of the Brillouin zone  (BZ).
The self-energy 
is adjusted  at each iteration until self-consistency is reached.
Since this method has not been used previously, we provide additional 
details for purposes of clarity in Appendix A, which provides a 
flow chart and a step-by-step description of the procedure.

\subsubsection{Density matrix self-consistency}
In a single site approximation, the local Green's function matrix $G(z)$ for
the  5$f$ electrons in the manifold is
\begin{eqnarray}
\label{eq:gf2}
G(z) =  \Big( {G}_{0}^{-1} + \Delta \mu - \Sigma(z) \Big)^{-1} \, ,
\end{eqnarray}
where  ${G}_{0}(z)$ is the non-interacting Green's function 
% (DFT, an analog of the static Weiss field in DFT+DMFT).
 $\Delta \mu$ is a correction to the chemical potential chosen to ensure that
$n_f = - \pi^{-1} {\rm Im} \; {\rm Tr}
  \int_{-\infty}^{E_{\rm{F}}} {\rm d}z  G(z)$
is equal to
the number of correlated $f$-electrons obtained from Eq.(\ref{eq:kohn_sham}).
Then, with the aid of this local Green's function $G(z)$, we evaluate
the occupation matrix
\begin{eqnarray}
n_{\gamma\gamma'} = -\frac1{\pi}\,\mathop{\rm Im}
\int_{-\infty}^{E_{\rm{F}}} {\rm d} z \, G_{\gamma \gamma'}(z).
\end{eqnarray}
For the energy integrations we use $\Im(z)/\pi=0.01$ eV, and a grid along
the real $z$ axis of 0.01 eV.

The matrix $n_{\gamma_1 \gamma_2}$ is used to construct an effective LDA+$U$
potential ${V}_{U}$, which is inserted into the Kohn--Sham-like
equations~\cite{shick99}:
\begin{gather}
\bigl[ -\nabla^{2} + V_{\rm LDA}(\mathbf{r}) + V_{U} + \xi ({\bf l} \cdot
{\bf s}) \bigr]  \Phi_{\bf k}({\bf r}) = \epsilon_{\bf k}^b \Phi_{\bf
k}({\bf r}).
\label{eq:kohn_sham}
\end{gather}
For the spherically-symmetric LDA+U  double-counting term (included in  
the potential $V_U$)  we have adopted the fully localized limit
(FLL) form  $V_{dc} = U (n_f-1/2) - J(n_f-1)/2$. We also note that the LDA  potential
$\hat{V}_{\rm LDA}$ in Eq.(\ref{eq:kohn_sham}) acting on the $f$-states 
is corrected to exclude  the non-spherical double-counting 
with $V_U$.\cite{Kristanovski2018}
%The DFT+U Green function matrix $G_{+U}$ is calculated from
%Eq.~(\ref{eq:gf}) substituting the self-energy $\Sigma(\epsilon)$ by 
%the DFT+U potential $V_{U}$. 
The equations in Eq.~(\ref{eq:kohn_sham}) are iteratively solved until
self-consistency over the charge density is reached.  
% The equations (3)-(7)
%  Eq.(\ref{eq:kohn_sham}) 
% are iteratively solved until self-consistency over
% the charge density and occupation matrix is reached.
The DFT+U Green function matrix $G_{+U}$ is calculated from Eq.~(\ref{eq:gf})
substituting the self-energy $\Sigma(\epsilon)$ by the DFT+U potential
$V_U$.

The new value of the 5$f$-shell occupation is obtained
from the solution of Eq.~(\ref{eq:kohn_sham}), and defines  
the new value of $\epsilon_f = -V_{dc}$ in 
Eq.~(\ref{eq:hamilt}).\cite{shick09}  
The $f$-shell SOC parameter (an atomic quantity), and the CF matrix 
$\Delta_{\rm CF}$ in Eq.~(\ref{eq:hamilt}) are determined
in each iteration. 
The CF matrix $\Delta_{\rm CF}$ in Eq.(\ref{eq:hamilt}) is obtained by 
projecting the self-consistent solutions of Eq.(\ref{eq:kohn_sham})  into
the $\{\phi_{\gamma}\}$ local $f$-shell basis, giving the ``local Hamiltonian"
\begin{eqnarray}
[H_{loc}]_{\gamma\gamma'} &=&
\int_{\epsilon_b}^{\epsilon_t} {\rm d} \epsilon \, \epsilon 
             [N(\epsilon)]_{\gamma \gamma'}   \nonumber \\ 
    &\approx& \epsilon_0 \delta_{\gamma \gamma'}
+ [\xi {\bf l}\cdot{\bf s} + \Delta_{\rm CF}]_{\gamma \gamma'}  + [{V_{U}}]_{\gamma \gamma'} \, ,
 \label{eq:hloc}
\end{eqnarray}
where $[N(\epsilon)]_{\gamma_1 \gamma_2}$ is the $f$-projected 
density of states (fDOS) matrix 
(whose integral to $E_F$ gives the familiar occupation
matrix), 
 $\epsilon_b$ is the bottom of the valence band, $\epsilon_t$ is the upper cut-off,
and $\epsilon_0$ is the mean position of the non-interacting $5f$
level. 
The matrix  $\Delta_{\rm CF}$ is then obtained by removing the interacting DFT+$U$
potential  and SOC  $[\xi {\bf l}\cdot{\bf s}]_{\gamma \gamma'}$ from 
$H_{loc}$ Eq.( \ref{eq:hloc}). As mentioned, 
for UTe$_2$ it is anticipated that the
CF is represented sufficiently by DFT+U, and this step is neglected.

The self-consistency loop is closed by calculating the non-interacting Green's function $G_0$,
\begin{eqnarray}
G_0(z) =  \Big( G_{+U}^{-1} + V_{U} \Big)^{-1}
\end{eqnarray}
and the next iteration is started by
solving~Eq.~(\ref{eq:hamilt}) for the updated $\epsilon_f$, 
$\xi$, and $\Delta_{CF}$,
and calculating the new self-energy $\Sigma(z)$.
The self-consistent procedure  was repeated until the
convergence of the 5$f$-manifold occupation matrix, with $n_f$ converged
to less than 0.01.

%%%%%%%%%%%%%%%%%%%%%%%%%%%%%%%%%%%%%%%%%%%%%%%%%%%%%%%%%%%%%%%%
\section{Computational Results}
\begin{figure*}[!htbp] 
%\vskip 8mm
%\centerline{$U$= 3 eV  \hspace*{9 cm} $U$=6 eV}
\centerline{\includegraphics[width=1.95\columnwidth]{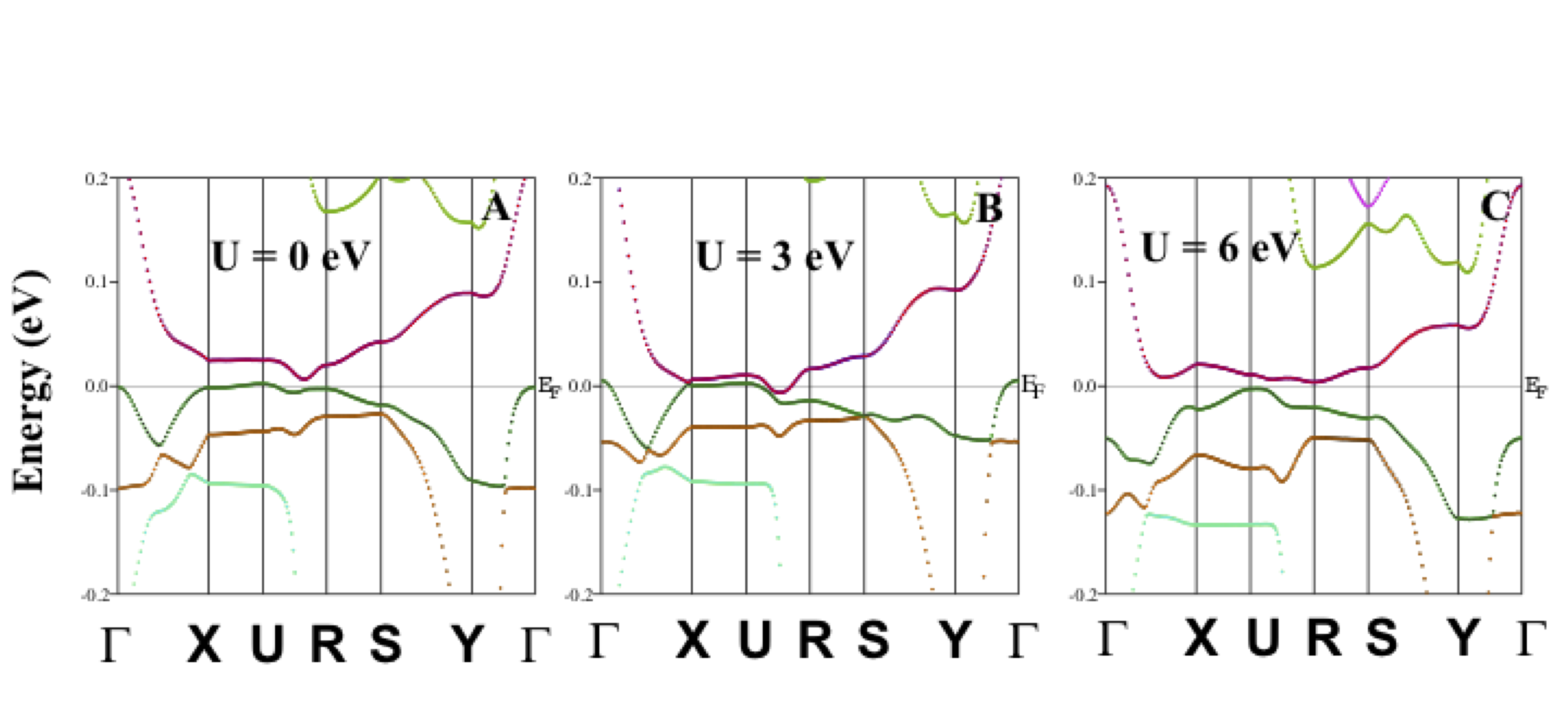}}
\caption{Band structures near the Fermi level $E_F$=0 for unpolarized 
UTe$_2$ from (A) the DFT+U(OP) functional, and from the DFT+U(ED)
functional for (B) $U$=3 eV, and (C) $U$=6 eV. Note that in (B)
for $U$=3 eV, a band crossing at E$_F$ lies very near the point $X$.
The special $k$-points lie along Cartesian directions: 
$\Gamma [0,0,0]$, X$ [\pi/a,0,0]$, U$[\pi/a,0,\pi/c]$, R$[\pi/a,\pi/b,\pi/c]$,
 S$[\pi/a,\pi/b,0]$, Y$[0,\pi/b,0]$.}
\label{fig:dos}
\end{figure*}

The U and Te atoms projected densities of states (DOS) for $U$=3 eV, 
and the total densities of states for $U$=3 eV and 6 eV, 
are shown in Fig.~\ref{str-hyb} The band structures for 
two different Coulomb~$U$ values are shown in Fig.~\ref{fig:dos}.  
For all values of $U$, the DOS near $E_F$ is almost entirely due to 
U 5$f$ states (see Table I). As several groups have noted, there is a
small gap for $U$=0, the curious LDA result. 

Increasing $U$, the two flat bands near E$_F$ 
become flatter, approach each other, and by $U$=3 eV become inverted,
leaving a band crossing along $\Gamma-X$ very near $X$ almost exactly
at the Fermi level.
Increasing $U$ to 6 eV, the bands separate leaving a small 10 meV bandgap, 
again separating disjoint valence and conduction bands. This moving away from
$E_F$ of the $5f$ bands is qualitatively consistent with DMFT results
using $U$=6 eV. However, we do not obtain any highly dispersive
Te $5p$, U $5d$ band in the background, crossing E$_F$, as in DMFT. In
fact, for $U$=6 eV we obtain again a small gap, whereas (to repeat)
UTe$_2$ is observed to be a heavy fermion metal, not
necessarily in conflict with a semimetal before dynamic correlations are
included. Henceforward we focus on our $U$=3 eV results with flat
$5f$ bands crossing $E_F$. 

\begin{comment}
Electronic structure theorists have been searching for an appropriate
band structure from which to address further question.
Similarly to previously reported LDA result just mentioned (a gap
incorrectly), Harima performed an {\it ad hoc} displacement of the
$5f$ states upward.\cite{Harima2019}  and DFT+U(OP) 
calculations~\cite{Shick2019},
UTe$_2$ is obtained as a narrow gap semiconductor ($U$=0 or $U$(ED)=6
eV) or a small overlap semimetal ($U$(ED)=3 eV. 
Experimentally, of course, UTe$_2$ is observed to be a HF metal, not
necessarily in conflict with a semimetal before dynamic correlations are
included, and we have focused on the $U$=3 eV results.  
\end{comment}

\subsection{Band structure and Fermi level quantities}
Density functional theory with (semi)local exchange-correlation functionals
($U$=0) give an insulating band structure for UTe$_2$, albeit with a very small
10-15 meV gap. Such a gap generally signals a bonding-antibonding
separation of bands, but no such description has been forthcoming for UTe$_2$. The 
gap reflects a different brand of band insulator. Every calculation
reveals that the large uranium SOC
separates the $j=\frac{5}{2}$ and $j=\frac{7}{2}$ subshells separated by
1.5 eV (see Fig.~\ref{str-hyb}), and with a $5f^3$
configuration the former subshell is half-filled. This SOC separation is
much larger than any crystal field splitting. The previous work of
some of the present authors\cite{Shick2019} established that, with
ferromagnetic order imposed, 
two orbitals, $m_j=\pm\frac{1}{2}$,
are half-filled, then hybridized, leading to the gap which is however
very small. How physical UTe$_2$ obliterates this gap and in the process
emerges as a nearly magnetic but superconducting material is
the fundamental issue in the electronic structure of this compound. 

Henceforward we focus on $U$=3 eV results unless otherwise stated, since
this value is sufficient 
to restore a conducting band structure, and 
is typical of values in most uranium intermetallics.
From Fig.~\ref{fig:dos}, this ``insulator-metal'' transition arises from
a hybridization reconfiguration 
of energy levels at the zone boundary points $U=(\frac{\pi}{a},0,\frac{\pi}{c})$
and $X=(\frac{\pi}{a},0,0)$.
A distinctive feature is that all four bands shown in Fig.~\ref{fig:dos}
are exceedingly flat along $X-U$ (the $k_z$ direction), 
unlike for $U$=0 or $U$=6 eV. Another
feature is the band crossing along $\Gamma-X$ very near the point $X$. 
The unoccupied band for $U$=6 eV is nearly dispersionless along the
three directions $X-U-R-S$, before mixing with dispersive bands in
other regions  of the zone.
 
The values of $N(E_F)$, $N_f(E_F$), and Fermi velocities 
along the three crystal axes for $U$=3 eV are provided in 
Table~\ref{tab:dos_velocities}. 
The  Fermi velocities  are the r.m.s. FS averaged values 
$v_{F,x}=\sqrt{<v_{k,xx}^2>_{FS}}$, and similarly for $yy$ and $zz$ components. 
The anisotropy is only 10-15\%, indicative of three dimensional conduction;
anisotropy is larger for the individual bands.
The magnitudes for the separate bands, 0.3-1.4 $\times 10^5$ cm/s, indicate very heavy
carriers even before renormalization by dynamical processes (electronic
and phononic).  

\begin{table}[htb]
%\begin{table}[floatfix]
\caption{The total $N(E_F)$ and $f$-projected $N_f(E_F)$ densities of states,
in eV${}^{-1}$,
and the direction-resolved Fermi velocities in units of 10$^5$ cm/s for $U=$ 3 eV.
The electric field gradients (EFG) $V_{xx}, V_{yy}, V_{zz}$ (subscripts denote
second derivatives) are in units of
10${}^{21}$ {V}/{m$^2$}, and the dimensionless asymmetry parameter $\eta$
is given. The two Te sites have EFGs differing by factors of 7-8.}
\label{tab:dos_velocities}
\begin{tabular}{|r|rr|rrr|}
  \hline
 &N($E_F$)&$N_f$($E_F$)& $v_{F,x}$ & $v_{F,y}$ & $v_{F,z}$  \\
  \hline
Total &11.04&10.01&0.88& 0.64 & 0.56 \\
FS-I  & 6.23 & 5.67 &0.48&0.72 & 0.32\\
FS-2 & 4.81 & 4.34 & 1.44& 0.80 &0.80 \\
  \hline
  \multicolumn{6}{c}{Electric Field Gradient} \\
  \hline
 Atom & site &  $\eta$ & $V_{zz}$ & $V_{yy}$ & $V_{xx}$ \\
 \hline
 U    & 4i   & 0.183 & 13.85 & -8.19 & -5.66 \\
 Te1  & 4j   & 0.173 &4.56   & -2.67 & -1.89 \\
 Te2  & 4h   & 0.176 &33.28 & -19.56 & -13.71 \\
 \hline
\end{tabular}
\end{table}

From Table~\ref{tab:dos_velocities} one sees that 90\% of $N(E_F)$ is
provided by the U 5$f$ states.
$N(E_F)$= 11.0 states/eV  for $U=3$~eV
corresponds to a band Sommerfeld constant $\gamma$=13.0 mJ/mol-K$^{-2}$. This implies
a mass enhancement of nearly 9 from dynamic interactions compared to the
experimental value. Note that there is a strong peak in the DOS (up to 59
states/eV) just  10 meV above $E_F$, corresponding to 
$\gamma$=70 mJ/mol-K$^{-2}$ which is within a factor of two of the experimental 
value of $\gamma$=120 mJ K$^{-2}$ mol$^{-1}$.\cite{Ran2019}
The narrow peak just above $E_F$ implies a strong dependence of properties on stoichiometry.

\begin{figure*}[!htbp]
\vskip -10mm
\centerline{\includegraphics[width=1.8\columnwidth]{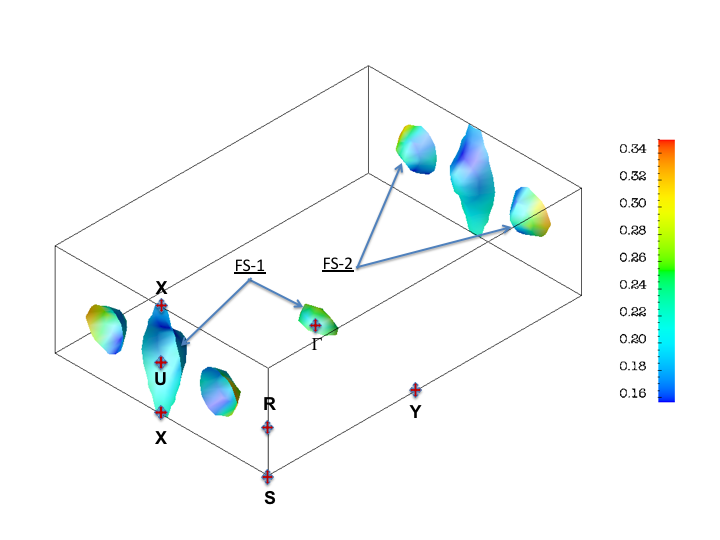}}
\centerline{\includegraphics[width=1.65\columnwidth]{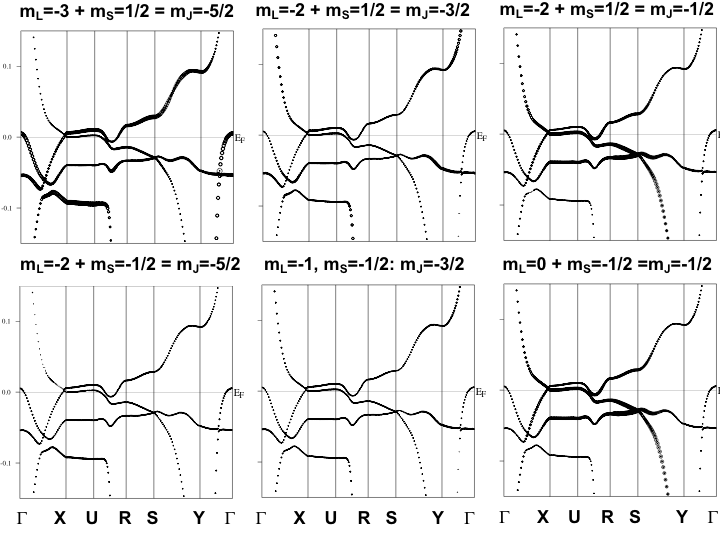}}
\caption{Upper panel:
The Fermi surfaces of UTe$_2$ from DFT+U(ED), for Coulomb $U$=3 eV. The colors
provide the relative Fermi velocities.
The high symmetry k-points are
$\Gamma [0,0,0]$, X$ [\pi/a,0,0]$, U $[\pi/a,0,\pi/c]$, R $[\pi/a,\pi/b,\pi/c]$,
 S $[\pi/a,\pi/b,0]$, Y $[0,\pi/b,0]$. Lower panels:
the $5f$ fat-band structure of UTe$_2$, with the circle size indicating 
the amount of $m_l, m_s, m_j=m_l+m_s$ character in the bands, as labeled.
}
\label{fig:fs-3V}
\end{figure*}

In materials with large SOC where the $j=\frac{5}{2}, j=\frac{7}{2}$ splitting
dominates site anisotropy and crystal field splitting but symmetry is low, 
state characters are not very transparent. 
The $|m_l,m_s >$ and $|j,m_j=m_l+m_s >$ decompositions of $N_f(E_F)$ 
are provided in Appendix B,  Table~\ref{dosef}.
The important bit of information is that the 
$|\frac{5}{2},\pm\frac{1}{2}>$
   components (equal by symmetry) are three times larger than the 
$|\frac{5}{2},\pm\frac{5}{2}>$ components, and five times larger than
$|\frac{5}{2},\pm\frac{3}{2}>$ components,
with the $m_j=\pm\frac{7}{2}$ components being negligible. This values 
reflect strong
spin-orbital polarization at the Fermi level in UTe$_2$, which also shows
up in the strong anisotropy of the electric field gradients, below.

The Fermi surfaces (FS)  are displayed in the upper panel of 
Fig.~\ref{fig:fs-3V}.
The FS has three types of sheets: from the lower band is the hole sheet centered
at $\Gamma$ and a fluted hole column along $X-U$, with masses varying 
by $\sim$50\% over the sheets.  The second band gives two symmetry related 
electron ellipsoids midway between $U$ and $R$, providing the required
charge compensation.
The corresponding Fermi velocities, with relative values shown by the
colorbar in Fig.~\ref{fig:fs-3V} and mean values provided in 
Table~\ref{tab:dos_velocities},
have somewhat less than factor-of-two anisotropies.

The array of band graphics in the lower part of Fig.~\ref{fig:fs-3V}, 
presented with fat-band character
and plotted along Cartesian directions, provides the relative 
amounts of the stated spin-orbital characters of bands near the Fermi level. 
The major contribution to FS-1 along the $X-U$ line arises from
$|\frac{5}{2},\pm\frac{1}{2}>$ orbitals. The $\Gamma$-point centered hole spheroid is 
more $|\frac{5}{2},-\frac{5}{2}>$ in character.
The electron sheet along $U-R$ arises from a mixture all three of these 
orbitals. 
%({\it i.e.} $|m_j=\pm\frac{1}{2}>$ and $|m_j=-\frac{5}{2}>$.
As in our previous work\cite{Shick2019}, we find that the Fermi level
states are dominated by $|\pm\frac{1}{2}>$ orbitals with some contribution
from the $|-\frac{5}{2}>$ orbital. 

Our DFT+U(ED) results, obtained without restriction to a
single-determinant reference state in determining the density,
can be contrasted with
previous beyond-DFT results. Admittedly, the groups that are involved
are searching for a treatment that will provide a realistic platform
for further considerations and experimental properties.  Conventional
DFT+$U$ was applied by Ishizuka {\it et al.}~\cite{Ishizuka2020}, who
chose the non-standard Hund's $J$=0 approach, neglecting Hund's
exchange and anisotropy of the $U_{m,m',m'',m'''}$ matrix. A crossover
in behavior was found for values of $U$ in the 1-1.5 eV range. Our earlier
treatment of orbital polarization by the DFT+U(OP)
method\cite{Shick2019} focused on ferromagnetic alignment based on the
observed large Curie-Weiss moment (which however does not order) so it
is less directly comparable. The problem posed by the unphysical gap in
LDA was addressed by Harima\cite{Harima2019} with a hands-on shift of
the U $5f$ energy by 1.36 eV, giving small Fermi surfaces arising from
flat bands much as we find, and by 2.72 eV, which led to a band structure
with large Fermi surfaces. As mentioned, the work of
Miao {\it et al.}\cite{Miao2019}
was strongly influenced by the apparent similarity of a dispersive band
crossing $E_F$ in ARPES that is like that in ThTe$_2$, which has no
$5f$ bands.

\subsection{Local U atom $f$-shell properties}
The calculated 5$f$ occupation within the uranium atomic sphere is $n_f$=2.73,
close to the value obtained in DFT+U(OP) calculations~\cite{Shick2019},
and supporting the viewpoint of a reference  U 5$f^3$ configuration. 
This value is a slight underestimate, since $5f$ orbitals extend somewhat
beyond the atomic spheres used to obtain this number. The electronic 
structure results discussed above have been more consistent with an
$f^2$ viewpoint, but without the comparison of calculated moments with
the experimental Curie-Weiss moment that we have provide below.

\subsubsection{X-ray absorption}
X-ray absorption spectroscopy (XAS) provides additional local 
information. XAS yields two intensities 
$I_{5/2}:4d_{5/2}\rightarrow 5f_{5/2,7/2}$ and
$I_{7/2}:4d_{3/2}\rightarrow 5f_{5/2}$, and the
branching ratio\cite{Moore} $B$=$I_{5/2}/(I_{5/2}+I_{3/2})$=0.71.
These are connected to the partial occupations
$n_f^{5/2}$ = 2.47 and $n_f^{7/2}$ = 0.26.
Our value of $B$ can be directly compared to future experimental 
results for XAS 
and electron energy-loss spectroscopies. The $|j,m_j>$ decompositions of
the U 5$f$-occupations are provided in Appendix B. 

\subsubsection{Curie-Weiss moment}
For the self-consistently determined impurity energy position 
$\epsilon_f = - V_{dc}$,
we obtain a doubly (Kramers) degenerate ground state with
spin, orbital, and total moments of $S$=-1.34, $L$=5.80, $J$=4.53, in $\mu_B$.
The calculated $g$-factor is 0.78, and $\langle m_j \rangle = \pm 0.55$.
The Curie-Weiss magnetic moment  $\mu_{eff} = g \sqrt{J(J+1)}$ = 3.52 $\mu_B$,
calculated as the rms value over occupied uranium configurations,
is in reasonable agreement with experimental values of 
2.8$\mu_B$~\cite{Ran2019}, and 3.3$\mu_B$~\cite{Aoki2019}. These can be
compared to the textbook values for an $f^3$ configuration:
$S=-\frac{3}{2}, L=6, J=\frac{9}{2}$, resulting in g=0.72. These Hund's
rule numbers, not normally reliable for $5f$ materials, are remarkably
similar to the DFT+U(ED) results. 

\subsubsection{Mass enhancement}
We obtain an estimated average mass enhancement, without contributions
from dynamical corrections, in two ways.
From the mean spectral density,
$${m^* \over m} =  {Tr[\hat{Z}^{-1} \hat{N}(E_F)] \over Tr[\hat{N}(E_F)]} \,,$$
where $\hat{Z}^{-1} = 
    [\hat{I} - d Re[\Sigma(\epsilon)]/d \epsilon]^{-1}$ 
is the quasiparticle residue matrix, and 
    $\hat{N}({E_F}) = - {1 \over {\pi}}\mathop{\rm Im}\mathop{\rm Tr} [G(E_F)]$
is the spectral density matrix obtained from Eq.(\ref{eq:gf}). 
The quasiparticle weight
$Z (= {m \over m^*}) = $ 0.06 is obtained. This small value of $Z$ 
indicates the strongly correlated character of $f$-electrons at $E_F$. 
The renormalized perturbation theory~\cite{Hewson} expression for the Kondo
temperature is
\begin{eqnarray} 
T_K = {\pi^2 \over 4} Z \Delta(E_F).
\end{eqnarray}
From the value of the hybridization function $\Delta(E_F)\approx$ 50 meV, 
the predicted $T_K \approx$ 100 K, very similar to
experimental values from resistivity.\cite{Ran2019}

\subsubsection{Electric field gradients}
Electric field gradients (EFG) provide a measure of the 
charge distribution (mostly from the local
charge) that is available from all-electron calculations.
Following the analysis of the electronic structure quantities given above, 
we have calculated the electric field gradients~\cite{Mohn2000} in UTe$_2$. 
Their values, together with the dimensionless asymmetry parameter 
for the $x-y$ plane values,
$$\eta = \frac{V_{xx} - V_{yy}}{V_{zz}}
  \equiv \frac{V_{yy}-V_{xx}}{V_{yy}+V_{xx}}$$
are provided in Table~\ref{tab:dos_velocities}. The second expression follows
from the traceless nature of the EFG tensor.
The EFG component $V_{zz}$ is proportional to the nuclear
quadrupolar resonance (NQR) frequency $\nu_Q$. Experimental 
measurements of NQR require stable isotopes with 
nuclear spin~$I\geq 1$. No such measurements have performed
to date, so our calculations provide a prediction of this measure of the
anisotropy of the charge density and resulting Hartree potential at the
nuclei.

The most notable result is that the values for the Te2 site are roughly
a factor of 7 larger than for the Te1 site, reflecting a substantially
different charge distribution around the two sites. The values for uranium
lie midway between, and the anisotropy factors $\eta\approx 0.18$ are
the same for all three atomic sites. Measurement of some of these will
provide useful information on the electronic density distribution,
and orbital polarization, of UTe$_2$.

\begin{figure}[!htbp]
%\vskip -5mm
%\centerline{\includegraphics[width=0.8\columnwidth]{fig4.png} }
\includegraphics[width=0.75\columnwidth]{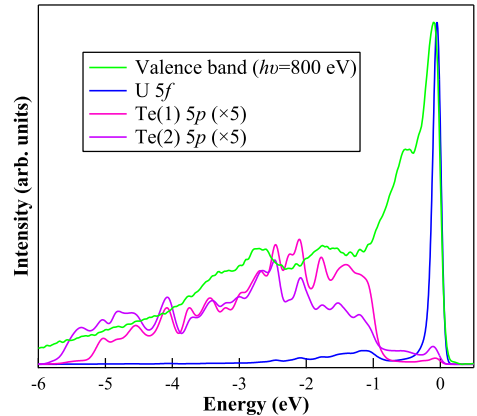}
%\vskip -4mm
\caption{Measured angle-integrated photoemission spectrum (green) 
plotted with the resolution broadened (120 meV Gaussian) and Fermi-Dirac cutoff (20 K)
DFT+U(ED) DOS.
The agreement of the onsets below E$_F$ reflect flat and heavy bands
that become the basis of the heavy fermion superconductor state.
The broadening of the experimental spectrum down to $\sim$-4 eV 
is the expected effect of
dynamical fluctuations in the uranium $5f$ shell. The
Te DOSs are enhanced by a factor of five to reveal possible Te
influence on structure in the -4 eV to -1 eV range, which is mostly
due to single particle excitations from the $5f$ shell.
}
\label{aipes}
\end{figure}

\section{Comparison between band structure and photoemission data}
Comparison of calculated bands (or spectral density) with 
PES data is the most direct means of determining the basis of the electronic
structure. Due to a number of experimental challenges --  matrix element
effects, energy and $k$ resolution, band broadening due to dynamical
effects, surface sensitivity -- comparison
can yet leave uncertainty, especially in quantum materials with 
strong dynamical processes.
Angle-integrated averages over the momentum and matrix element dependence,
giving the zone-averaged spectral density,
provides the most unambiguous information -- the spectral distribution of
valence band states -- on fundamental aspects of the
electronic structure. We remind that a band picture provides an optimum 
set of single particle characteristics (orbitals and eigenvalues) to best
describe ground state characteristics -- energies, charge and spin
densities and quantities derivable from them, for example, EFGs -- and
by continuity in metals, near ground state quantities. Single
particle excitations involve self-energies that are minor in many metals
but become central in describing heavy fermion metals.
Occupied multiplets and configurations are sampled in the DFT+U(ED) method
to determine spin-orbital occupations, while dynamic effects are not 
included in the band structure we present.  

\begin{figure*}[!htbp]
%\centerline{\includegraphics[width=1.75\columnwidth]{fig5.png} }
\includegraphics[width=1.75\columnwidth]{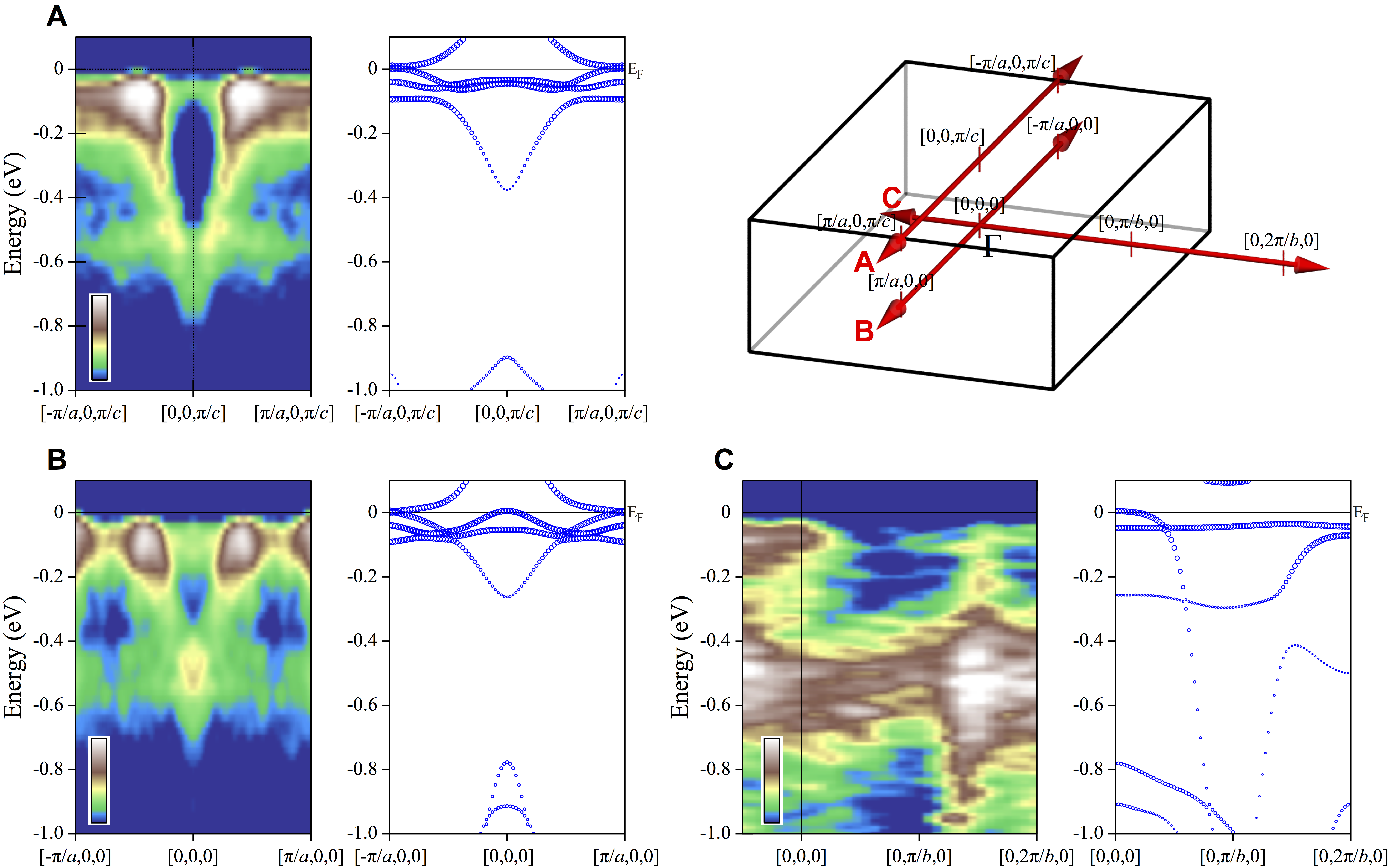}
\caption{
Upper right panel: Brillouin zone structure of UTe$_2$. panels A, B, C:
ARPES spectra (left) together with the band structure (right)
for the designated directions
and with energies aligned. The size of the calculated points provide the
relative amount of $5f$ character of the wavefunction. 
Panels A and C provide new data, while panel B
is a replot of earlier data\cite{Fujimori2019} of one of the authors
to provide clear comparison with our theoretical results.
The  k-points notations of the orthorhombic pseudo-zone
$\Gamma [0,0,0]$, X$[\pi/a,0,0]$, U $[\pi/a,0,\pi/c]$, Z $[0,0,\pi/c]$,
 Y$[0,\pi/b,0]$, $\Gamma' [0,2\pi/b,0]$ are used in the text. 
}
\label{fig:bands}
\end{figure*}

\subsection{AIPES}
Data taken at 20K with $h\nu$=800 eV
photon energy and 120 meV resolution are compared in Fig.~\ref{aipes}
with the resolution-broadened %and cross-section  weighted \cite{Fujimori2019} 
DFT+U(ED) DOS; other details of the experiment and analysis are described by
Fujimori and collaborators.\cite{Fujimori2019}. The unbroadened DOS is
shown in Fig.~\ref{str-hyb}.
As mentioned above, the most significant region for comparison 
is at and close below the Fermi energy,
where the dynamical self-energy is small. The leading edges at low energy 
in Fig.~\ref{aipes} are in extremely good experiment-theory
agreement. This result is crucial, because DFT+U(ED)
displays flat $5f$ bands at and immediately below E$_F$, whereas both
DFT+$U$\cite{Ishizuka2020} and DFT+DMFT\cite{Miao2019} displace $5f$ weight 
away from E$_F$ by several hundred meV for DMFT, or more for DFT+$U$. 
This energy shift is much larger than the experimental 
resolution, so the corresponding ``$5f$ edge'' in those spectra should be displaced 
by this amount from E$_F$, which is not seen in the data. 
These AIPES data thus support the view that 
flat bands lie at E$_F$, consistent with 
giving small renormalized Fermi surfaces that provide the platform for
the observed exotic superconducting states that are observed to be
extraordinarily sensitive to magnetic field, pressure, and stoichiometry. 

The AIPES intensity extending to 3-4 eV binding energy can be interpreted
in terms of the excitations involving $5f^3$ and $5f^2$ multiplets, as
discussed by Miao {\it et al.}\cite{Miao2019} These considerations involve
the relative participations of populations $f^2$ and $f^3$. The
DFT+DMFT treatment of Miao {\it et al.}
lead to a dominant $f^2$ ($^3H_4$, 84\%) state; DFT+$U$ (with $U$ and $J$
values described above) gives exactly $f^2$ (two
strongly bound $5f$ bands). Our DFT+$U$(ED) method leads to a dominant $f^3$
description. Summing the spin-orbital occupations provided in Appendix B 
(2.73) gives a slight underestimate due to $5f$ orbital tails extending beyond
the sphere boundary. The disjoint valence and conduction bands discussed
above argues for a half-filled $j=\frac{5}{2}$ subshell, which is $f^3$.

\subsection{ARPES}
 
 ARPES data were obtained in the
photon energy range 565-800 eV with energy resolution of 90-115 meV, 
with other aspects of the sample and setup described in Ref.~\cite{Fujimori2019}.
In Fig.~\ref{fig:bands} we show ARPES results compared with the relevant
band lines. 
For the U-Z-U line shown in Fig.~\ref{fig:bands}A,
agreement between the correlated bands and ARPES data near $E_F$ is  
apparent, with  a dispersive band dropping down at $Z=[0,0,\pi/c]$ 
being resolved clearly.
For the X-$\Gamma$-X direction 
in Fig.~\ref{fig:bands}B,
the calculations yield heavy $f$-bands located near and touching $E_F$. 
The bands lie at the same
energy as intense emission in the experimental data, and are separated by less than
the experimental resolution and sometimes crossing, so theory and experiment
are consistent although incoherence in the data at this low energy cannot be ruled out.  
This intensity is at variance with the presence of a light band observed by 
Miao {\it et al.}\cite{Miao2019} using photon energies in the 30-150 eV range.

Along the $\Gamma$-Y-$\Gamma'$ line in Fig.~\ref{fig:bands}C, incoherence in the
data lies in the energy region of a dispersive band, apparently
reflecting weakly dispersing, largely incoherent, $5f$ shell excitations.
The strong intensity at [0,0,0] around -60 meV is in a region where
our band structure predicts flat $5f$ bands. Note also that Fig.~\ref{fig:bands}C
indicates a dispersive Te $p$ band passing through the $5f$ bands and mixing
strongly in the calculations. Thus $f-p$ mixing is substantial, with a result
that the Te $p$ character is strongly excluded from the $5f$ band regions, as
is clear from the projected DOS in Fig.~\ref{str-hyb}. 

Both the calculated bands near E$_F$ as well as 
this data, and also the ARPES data of Ref.~[\onlinecite{Fujimori2019}], differ
from the results of Ref.~[\onlinecite{Miao2019}].
We attribute the differences in the experimental ARPES data 
%  of  Ref.~\cite{Fujimori2019} and  Ref.~\cite{Miao2019}
to the higher surface sensitivity of the spectra taken in the 30-150 eV
\cite{Miao2019} where the escape depth is $\sim$8-12\AA.  
The energy dependence of the electron escape 
depth\cite{damascelli2004} indicates that it
is roughly twice as large\cite{damascelli2004} in our energy range.
Thus our data are more bulk sensitive while
those of Miao {\it et al.} are more surface impacted,
where confinement imposed by the surface can lead to the U 5$f$-electrons 
becoming more localized than in the bulk. 
The low energy ARPES data of Miao {\it et al.}
were mostly interpreted by the band structure of ThTe$_2$, {\it i.e.}
any without $5f$ bands whatsoever. 
Conversely, our ARPES data together with
our correlated band results, along with earlier AIPES data,\cite{Fujimori2019} 
emphasize the presence of heavy 5$f$ bands near $E_F$.

\section{Summary}
While a great deal of experimental data has been collected that is 
relevant to the complex phase
diagram of UTe$_2$, and there are several theoretical suggestions 
about the character
and symmetry(s) of its superconducting and magnetic phases, there
is not yet any consensus emerging on its basic electronic structure.
Given its heavy fermion properties this may not be so surprising, but relative
to modestly correlated DFT-based calculations, the dynamical mass
enhancement is substantial but not particularly large. 
The Hill criterion for the critical uranium atom 
separation\cite{Hill1970,Boring2000} is 3.5~\AA, and clearly UTe$_2$ lies
on the localized side of that limit, but it does not order magnetically at
zero field. The Hill criterion is however
sometimes violated in uranium compounds, so th guidance it provides is
limited.
 
There is evidence that neither a fully localized nor simple itinerant 
picture holds for UTe$_2$.
Our correlated band DFT+U(ED) calculations, motivated by the Anderson impurity
model and taking into account the multiconfiguration aspect of 
the U $5f$ shell, 
suggest that both local and itinerant characteristics
of the $f$-electrons appear near the Fermi level and need to be treated
together. The uranium  magnetic moment is given well compared to the two
experimental reports, and an unexpected result is that the spin, orbital,
and total moments are near the Hund's rule prediction for an $f^3$ ion, and
not representative of an $f^2$ ion. 

The simplest and clearest experimental information on the electronic structure
is from angle-integrated PES. The occupied $5f$ spectral 
density\cite{Fujimori2019} 
peaks immediately below the Fermi level, and is strong within
1 eV of E$_F$ but contains structure (presumably) satellites 
out to 4 eV binding energy. The measured spectrum
spectrum represents broadening of a dynamical origin that
our method does not take into account. The other two means of including
Coulomb repulsion $U$, DFT+$U$ and DFT+DMFT, both displace the $5f$ spectral
density to higher binding energy, leaving only a $\delta$-function-like
Kondo peak at E$_F$ (DMFT), not consistent with AIPES data. Our spin-orbital
occupations also support dominance of a $f^3$ ion.

We have compared the DFT+U(ED) bands with the ARPES spectra of Fujimori
and collaborators\cite{Fujimori2019} and new data, which used higher 
photon energy with a
larger escape depth and thus more bulk sensitivity than the data of
Miao {\it et al.}\cite{Miao2019} The comparison at low binding energy
is encouraging, especially
considering some uncertainty in extracting bulk (three-dimensional) band
information from emission of electrons through a single surface. For these
reasons, we propose that our DFT+U(ED) bands provide an appropriate basis
of understanding and building on the electronic structure. We further comment
that the flat $5f$ bands crossing E$_F$ lead to small Fermi surfaces that
will be responsive to pressure and magnetic field, a sensitivity that is
very clear in the emerging experimental phase diagram. 

Finally, our low energy band structure shows evidence of Te $6p$ -- U $5f$
mixing, which repels the Te $6p$ character from the low energy region 
rather than opening up to allow a dispersive Te band to cross the
Fermi level. Still,
the Te influence is important, and the large U-U separation placing it in
magnetic ion regime suggests that exchange coupling in UTe$_2$ proceeds 
primarily of RKKY character mediated through the Te conduction bands. 
The overall picture is one of
Fermi level bands dominated by $5f$ character as in UPd$_3$ versus a
more strongly localized moment as in the Kondo lattice picture.

\section{Acknowledgments}
We thank G. R. Stewart for guiding us to the updated uranium Hill plot
published in Ref.~[\onlinecite{Boring2000}].
A.B.S. acknowledges partial support provided by Operational Programme Research, Development and Education financed by European Structural and Investment Funds and the Czech Ministry of Education, Youth, and Sports (Project No. SOLID21 - CZ.02.1.01/0.0/0.0/16$_{-}$019/0000760), and by the Czech Science Foundation (GACR) Grant No. 18-06240S.
The experiment was performed under Proposal No.2019A3811 at SPring-8 BL23SU. 
S.-i.F. was financially supported by JSPS KAKENHI Grant Numbers JP16H01084 and No. JP18K03553.
W.E.P. was supported by National Science Foundation Grant DMR 1607139.
\clearpage

\appendix
\section{Density matrix self-consistency in DFT+U(ED)}
\begin{figure}[!ht]
\includegraphics[width=1.0\columnwidth]{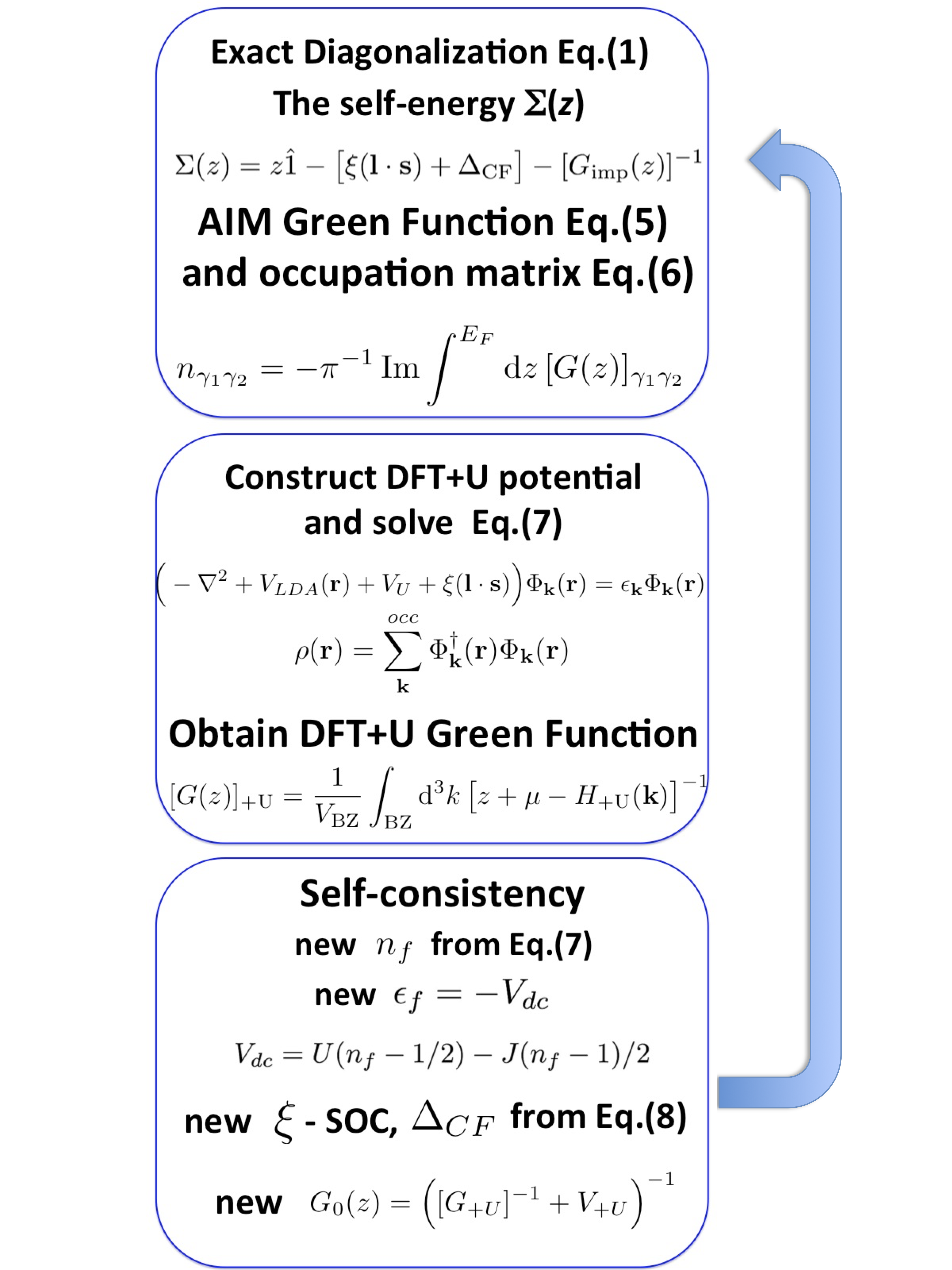}
\caption{Flow chart indicating the main steps in the charge and
occupation matrix self-consistent procedure for our DFT+U(ED)
implementation. See the description in Appendix A.
}
\label{flow}
\end{figure}

%%%% XXXX
The calculation follows in several respects that of Shick and
collaborators\cite{shick09} for their ``DFT+Hubbard I approximation''
study of elemental actinides. A few generalizations have been
adopted here, with the main steps being illustrated in the flow diagram
in Fig.~\ref{flow}.
The various steps occur in the following order.
\begin{enumerate}
\item From a DFT/LDA calculation the crystal Green's function $G_0(z)$ is constructed
and initial parameters to solve Eq.(1) are chosen as reasonable estimates.
For example, the repulsion $U$ and Hund's rule $J$ constants (from
the Slater parameters) must be
specified.
\item From the (self-consistent) DFT calculation, 
the hybridization matrix
is evaluated, from which the bath energies $\epsilon^j_{q=1}$ and
hybridization strengths $V_q^j$ are
chosen to represent mixing with the environment, as described in Sec. II.C.
The input parameters to the DFT+U(ED) calculation are now determined.
\item Carry out an exact diagonalization of $H_{int}$ of Eq. (1).
This step is represented by the top panel of the flow chart
in Fig.~\ref{flow}. Use 
eigenvalues and eigenvectors to construct the impurity Green's
function $G_{imp}$ in Eq.~(\ref{eq:gimp}) and the corresponding self-energy;
adjust $\mu$ in Eq.~(\ref{eq:gf2}) to fix the value of $n_f$ to the current
value from the full calculation equation~(\ref{eq:kohn_sham}).  
Calculate the occupation matrix equation~(6) as stated in Sec. II.C.3. 
%See the top panel in Fig.~\ref{flow}. 
%
\item From the occupation matrix, set up the DFT+U potential, 
solve self-consistently
the Kohn-Sham equation~(\ref{eq:kohn_sham}), to find the updated 
charge density according to some reliable prescription typically
present in DFT codes. Obtain the DFT+U Green's function.
See the central panel in Fig.~\ref{flow}.
\item Obtain a new non-interacting Green's function $G_0$, a 
new position $\epsilon_f$ 
of the impurity level, a new $\xi$ and (when used) $\Delta_{CF}$. 
This step is represented in the bottom panel in  Fig.~\ref{flow}.
\item
Now close the self-consistency loop. 
When output and input of the $5f$-manifold occupation $n_f$ convergence agree within 
a specified criterion, the self-consistency
loop is exited, and analysis of the results follows.
\end{enumerate}

\vskip 8mm
\section{Decomposition of the U 5$f$ DOS at $E_F$, and the U 5$f$ occupations}
The essential aspects of the electronic structure of UTe$_2$
finally reduce to participation of various spin-orbitals in the $5f$
occupation, and to their play in states at the Fermi level. Table
~\ref{Fermi-level} provides the Fermi level quantities in both ($\ell$,s)
and ($j,m_j$) representations. The primary result to notice is the large
and equal participation at E$_F$ of the $m_j=\pm\frac{1}{2}$ orbitals.

\begin{table}[!htbp]
%\vspace*{-0.25cm}
\caption{The $|m_l, m_s >$ and $|j,m_j>$ decompositions of
the U atom $f$-projected DOS at $E_F$ (in 1/eV) are provided 
for the unpolarized system. 
The magnetic quantization is along the easy $\hat{a}$ axis,
these values are for Coulomb $U$=3 eV. }
\label{dosef}
\begin{ruledtabular}
\begin{tabular}{ccccccccc}
% \multicolumn{8}{c}{$U$=3 eV} \\
$m_l$  & -3  & -2 & -1 & 0 & 1 & 2 & 3 \\
\hline
Spin-$\uparrow$     & 0.48 & 0.23 & 0.85 & 0.81 & 0.10 & 0.07& 0.01 \\
Spin-$\downarrow$& 0.01 & 0.07 & 0.10 & 0.81 & 0.85 & 0.23 & 0.48 \\
$m_j$  & -7/2  & -5/2 & -3/2 & -1/2 & 1/2 & 3/2 & 5/2 & 7/2 \\
            & 0.01  & 0.55 & 0.33 & 1.66 & 1.66 & 0.33 & 0.55 & 0.01 \\
\end{tabular}
\end{ruledtabular}
\label{Fermi-level}
\end{table}

Table~\ref{five-halves} provides the strong spin-orbit decompositions
for moment along the easy $\hat a$-axis. The $j=\frac{7}{2}$ contribution
is minor.

\begin{table}[!htbp]
%\vspace*{-0.25cm}
\caption{The  $|j,m_j>$ decompositions of
the U atom $f$-occupation 
for the unpolarized system. 
The magnetic quantization is along the easy $\hat{a}$ axis.
Coulomb $U$= 3 eV and 6 eV values are provided. }
\label{occup}
\begin{ruledtabular}
\begin{tabular}{lcccccccc}
 \multicolumn{9}{c}{$U$ = 3 eV} \\
$\; j \;$ / $\; \; m_j$&-7/2 & -5/2 & -3/2 & -1/2 & 1/2 & 3/2 & 5/2 &7/2\\
 5/2                     &       & 0.412 & 0.410 & 0.415 & 0.415 & 0.410 & 0.412   \\
%$j=7/2, m_j$  & -7/2  & -5/2 & -3/2 & -1/2 & 1/2 & 3/2 & 5/2 & 7/2 \\
 7/2           & 0.031  & 0.037 & 0.031 & 0.030 & 0.030 & 0.031 & 0.037 & 0.031\\
 \multicolumn{9}{c}{$U$= 6 eV} \\
%5/2, m_j$  & -5/2 & -3/2 & -1/2 & 1/2 & 3/2 & 5/2  \\
 5/2                           &      & 0.410 & 0.417 & 0.410 & 0.410 & 0.417 & 0.410  \\
%$j=7/2, m_j$  & -7/2  & -5/2 & -3/2 & -1/2 & 1/2 & 3/2 & 5/2 & 7/2 \\
 7/2           & 0.027  & 0.019 & 0.033 & 0.022 & 0.022 & 0.033 & 0.019 & 0.027 \\
\end{tabular}
\end{ruledtabular}
\label{five-halves}
\end{table}

%\begin{comment}
%\bibliographystyle{apsrev}
%\bibliography{refs}
%\clearpage

\end{document}